\documentclass[journal]{IEEEtran}

\ifCLASSINFOpdf
\else
\fi

\usepackage[absolute]{textpos}
\usepackage{graphicx}
\usepackage{caption}
\usepackage[margin=0.64in,top=1in]{geometry}
\usepackage{amsmath, amssymb, bm, cite, epsfig}
\usepackage{epstopdf}
\usepackage{graphicx}
\usepackage{array}
\usepackage{multirow}

\usepackage{amsfonts}
\usepackage{tabularx} 
\usepackage{tabu}
\usepackage{pbox}
\usepackage{makecell}

\usepackage{booktabs}
\usepackage{ctable}
\usepackage{xcolor,colortbl}
\definecolor{Gray}{gray}{0.85}
\definecolor{LightCyan}{rgb}{0.8,1,1}

\def\beq{\begin{equation}}
\def\eeq{\end{equation}}
\def\beqa{\begin{eqnarray}}
\def\eeqa{\end{eqnarray}}
\def\beqan{\begin{eqnarray*}}
\def\eeqan{\end{eqnarray*}}

\def\PL{\mathrm{PL}}
\def\dB{\mathrm{dB}}

\def\tm1{t\! - \! 1}
\def\tp1{t\! + \! 1}

\def\PL{\mathrm{PL}}
\def\dB{\mathrm{dB}}

\def\FSPL{\mathrm{FSPL}}
\def\log{\mathrm{log}}

\def\CI{\mathrm{CI}}

\def\PL{\mathrm{PL}}
\def\dB{\mathrm{dB}}

\def\FSPL{\mathrm{FSPL}}
\def\log{\mathrm{log}}

\def\CI{\mathrm{CI}}

\def\m{\mathrm{m}}
\def\AT{\mathrm{AT}}

\renewcommand\IEEEkeywordsname{Index Terms}

\begin{document}
		\begin{textblock}{18.8}(1.5,0.5)
			\centering
			\noindent\Large S. Sun, G. R. MacCartney Jr., and T. S. Rappaport, "A novel millimeter-wave channel simulator and applications for 5G wireless communications," \textit{2017 IEEE International Conference on Communications (ICC)}, Paris, May 2017.
		\end{textblock}
\title{A Novel Millimeter-Wave Channel Simulator and Applications for 5G Wireless Communications}

\author{\IEEEauthorblockN{Shu Sun, George R. MacCartney Jr., and Theodore S. Rappaport}\\
\IEEEauthorblockA{NYU WIRELESS and NYU Tandon School of Engineering, New York University, Brooklyn, NY, USA 11201\\
\{ss7152,gmac,tsr\}@nyu.edu }

\thanks{Sponsorship for this work was provided by the NYU WIRELESS Industrial Affiliates program and NSF research grants 1320472, 1302336, and 1555332.}
}
\maketitle

\begin{abstract}
This paper presents details and applications of a novel channel simulation software named NYUSIM, which can be used to generate realistic temporal and spatial channel responses to support realistic physical- and link-layer simulations and design for fifth-generation (5G) cellular communications. NYUSIM is built upon the statistical spatial channel model for broadband millimeter-wave (mmWave) wireless communication systems developed by researchers at New York University (NYU). The simulator is applicable for a wide range of carrier frequencies (500 MHz to 100 GHz), radio frequency (RF) bandwidths (0 to 800 MHz), antenna beamwidths (7$^\circ$ to 360$^\circ$ for azimuth and 7$^\circ$ to 45$^\circ$ for elevation), and operating scenarios (urban microcell, urban macrocell, and rural macrocell), and also incorporates multiple-input multiple-output (MIMO) antenna arrays at the transmitter and receiver. This paper also provides examples to demonstrate how to use NYUSIM for analyzing MIMO channel conditions and spectral efficiencies, which show that NYUSIM is an alternative and more realistic channel model compared to the 3rd Generation Partnership Project (3GPP) and other channel models for mmWave bands. 
\end{abstract}


\IEEEpeerreviewmaketitle
\section{Introduction}
The construction and implementation of channel models are becoming increasingly important for wireless communication system design, and computer-aided design tools such as channel simulators are essential for performance evaluation of communications systems and for simulating network deployments, before moving forward with new technologies.

There are several channel simulators that have been developed and used by previous researchers~\cite{Seidel,Jae14,Yu15,Rap91_SIRCIM,SMRCIM,Raj11,Fung93}. For instance, Smith~\cite{Smith75} built simulation software for indoor and outdoor propagation channels by making use of the two-ray Rayleigh fading channel model developed by Clarke~\cite{Clarke68}. Fraunhofer Heinrich Hertz Institute developed a 3-D multi-cell channel model that can accurately predict the performance for an urban macrocell setup with commercial high-gain antennas, upon which a channel simulator has been built that supports features such as time evolution, scenario transitions, and so on~\cite{Jae14}. A channel simulator for indoor scenarios was developed  for machine-to-machine applications~\cite{Yu15}. Rappaport and Seidel developed a measurement-based statistical indoor channel model named SIRCIM (Simulation of Indoor Radio Channel Impulse Response Models) for the early development of WiFi ~\cite{Rap91_SIRCIM} and the corresponding simulation software to generate channel impulse responses (CIRs) for indoor channels operating from 10 MHz to 60 GHz. A similar open-source RF propagation simulator is SMRCIM (Simulation of Mobile Radio Channel Impulse Response Models), that was useful for simulating outdoor channels~\cite{SMRCIM,Raj11}. Another software simulation program, called BERSIM~\cite{Fung93}, developed by Fung \textit{et al.}, was able to simulate mobile radio communication links and calculate average bit error rate (BER) and bit-by-bit error patterns, that was useful for evaluating link quality in real time without requiring any radio frequency hardware.  

In this paper, we introduce an open-source channel simulator named NYUSIM~\cite{NYUSIM_UM}, which has been developed based on extensive real-world wideband propagation channel measurements at multiple millimeter-wave (mmWave) frequencies from 28 to 73 GHz in various outdoor environments in urban microcell (UMi), urban macrocell (UMa), and rural macrocell (RMa) environments~\cite{Rap13:Access,Rap15:TCOM,Samimi15:MTT,Samimi16:EuCAP,Sun16:TVT,Sun15:GC,Mac15:Indoor,ACM,Mac17}. NYUSIM provides an accurate rendering of actual channel impulse responses in both time and space, as well as realistic signal levels that were measured, and is applicable for a wide range of carrier frequencies from 500 MHz to 100 GHz, and RF bandwidths from 0 to 800 MHz. As of early 2017, over 7,000 downloads of NYUSIM have been recorded. The source code was written in MATLAB~\cite{Matlab} and a platform-independent graphical user interface (GUI) was created to facilitate the use of NYUSIM on machines using either Windows or Macintosh operating systems even without MATLAB installed. 


\section{Channel Model Supported by NYUSIM}
Omnidirectional channel models have widely been studied and adopted by industry and researchers around the world to assist in wireless system design, yet directional channel models are also important to properly design and implement antenna arrays to exploit spatial diversity and/or beamforming gain in multiple-input multiple-output (MIMO) systems~\cite{Ertel,Sun14}. NYUSIM generates sample functions of the temporal and spatial CIRs from both omnidirectional and directional channel models that are borne out by measurements and models of NYU WIRELESS~\cite{NYUSIM_UM,Rap13:Access,Rap15:TCOM,Samimi15:MTT,Samimi16:EuCAP,Sun16:TVT,Sun15:GC,Mac15:Indoor,ACM,Mac17}. This section provides a brief overview of the path loss model and clustering algorithm used in NYUSIM.

\subsection{Path Loss Model}
The close-in free space reference distance (CI) path loss model with a 1 m reference distance~\cite{Sun16:TVT,Mac17,Rap15:TCOM,Mac15:Indoor}, and an extra attenuation term due to various atmospheric conditions~\cite{Lie93}, is employed in NYUSIM, which is expressed as~\cite{Rap15,Rap15:TCOM,Sun16:TVT}:
\begin{equation}\label{CI1}
\begin{split}
\PL^{\CI}(f, d)[\dB]=&~\FSPL(f, 1~\m)[\dB]+10n\log_{10}\left(d\right)\\
&+\AT[\dB]+\chi_{\sigma}^{\CI} \text{,}\\
&\text{ where } d\geq 1~\m
\end{split}
\end{equation}

\noindent where $f$ denotes the carrier frequency in GHz, $d$ is the 3D T-R separation distance, $n$ represents the path loss exponent (PLE), $\AT$ is the attenuation term induced by the atmosphere, $\chi_{\sigma}^{\CI}$ is a zero-mean Gaussian random variable with a standard deviation $\sigma$ in dB, and $\FSPL(f, 1~\m)$ denotes the free space path loss in dB at a T-R separation distance of $1~\m$ at the carrier frequency $f$:
\begin{equation}\label{FSPL}
\begin{split}
\FSPL(f, 1~\m)[\dB]&=20\log_{10}\left(\frac{4\pi f\times 10^9}{c}\right)\\
&=32.4[\dB]+20\log_{10}(f)
\end{split}
\end{equation}

\noindent where $c$ is the speed of light in a vacuum, and $f$ is in GHz. The term $\AT$ is characterized by:
\begin{equation}\label{AT}
\AT[\dB]=\alpha[\dB/\m]\times d[\m]
\end{equation}

\noindent where $\alpha$ is the attenuation factor in $\dB/\m$ for the frequency range of 1 GHz to 100 GHz, which includes the collective attenuation effects of dry air (including oxygen), water vapor, rain, and haze~\cite{Lie93}. The parameter $d$ is the 3D T-R separation distance in~\eqref{CI1}.

\begin{figure}
\centering
 \includegraphics[width=3.3in]{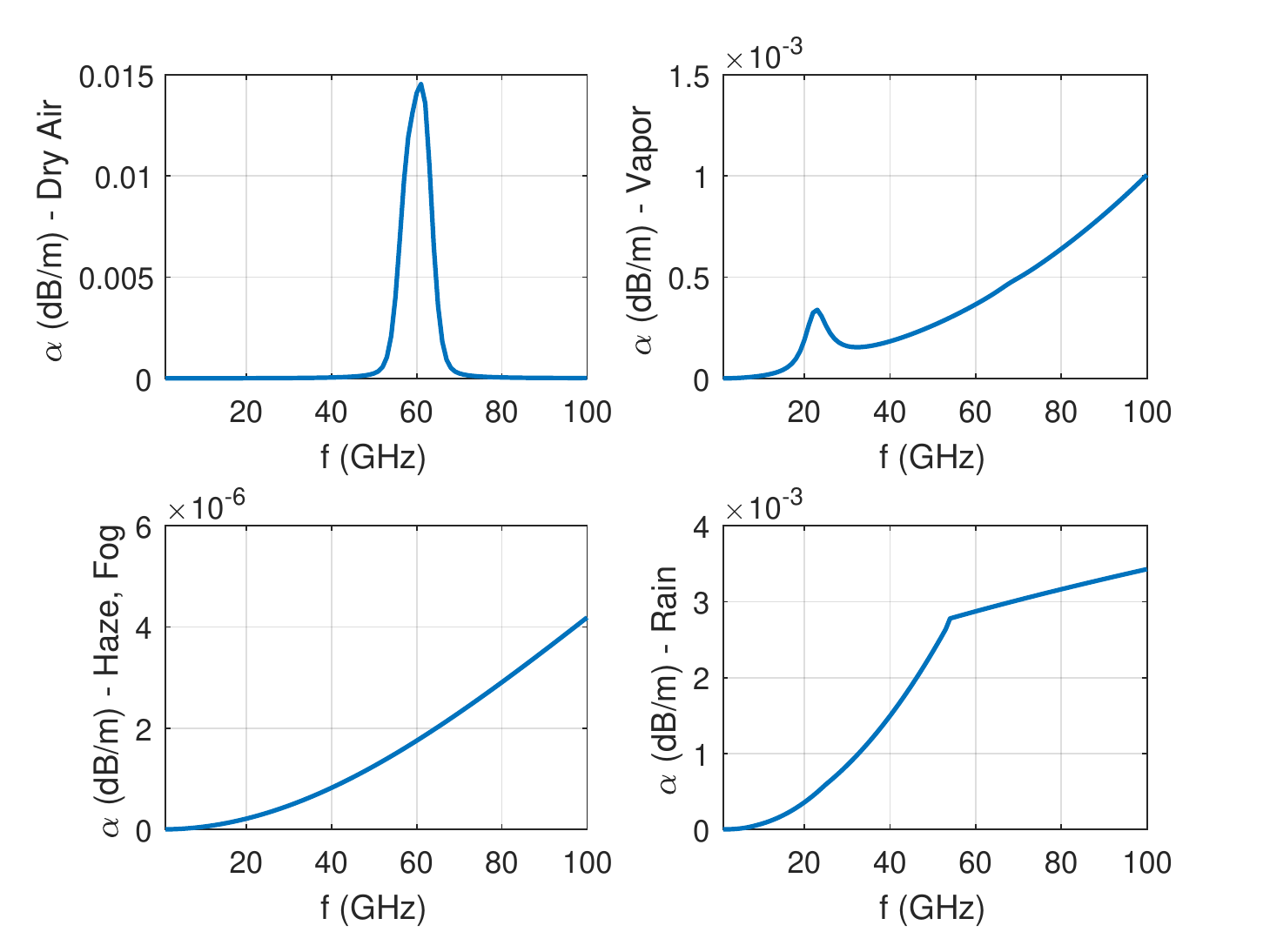}
    \caption{Propagation attenuation due to dry air, vapor, haze, and rain at mmWave frequencies, with a barometric pressure of 1013.25 mbar, a relative humidity of 80\%, a temperature of 20$^\circ$C, and a rain rate of 5 mm/hr~\cite{Lie93}.}
    \label{fig:AT1}
\end{figure}

Fig.~\ref{fig:AT1} illustrates examples of propagation attenuation values due to dry air, vapor, haze, and rain at mmWave frequencies from 1 GHz to 100 GHz, with a barometric pressure of 1013.25 mbar, a relative humidity of 80\%, a temperature of 20$^\circ$C, and a rain rate of 5 mm/hr. These results were obtained and reproduced from five reported controlled experiments on atmospheric attenuation~\cite{Lie93}.

Note that the CI model inherently has an intrinsic frequency dependency of path loss already embedded within the FSPL term, making it useful from below 1 GHz to above 100 GHz~\cite{Sun16:TVT,Mac17,Rap15:TCOM,Mac15:Indoor}. A useful property of~\eqref{CI1} is that 10$n$ describes path loss in dB in terms of decades of distances beginning at $1~\m$ (making it very easy to compute power over distance in one's mind). The CI path loss model is based on fundamental principles of wireless propagation, dating back to Friis and Bullington, where the PLE parameter offers insight into path loss based on the environment, having a PLE value of 2 in free space (as shown by Friis) and a value of 4 for the asymptotic two-ray ground bounce propagation model (as shown by Bullington~\cite{Bullington47a,Friis46a}). Hata adopted these fundamentals in his famous path loss model for the very early days of cellular~\cite{Hata,Andersen95}. Standardizing to a reference distance of 1 m makes comparisons of measurements and models simple, and provides a universal definition for the PLE, while enabling intuition and rapid computation of path loss~\cite{Sun16:TVT}. Compared with the existing alpha-beta-gamma (ABG) path loss model used in 3GPP/ITU channel models~\cite{Mac15:Indoor,Sun16:TVT}, the CI model uses fewer parameters while offering intuitive physical appeal, better model parameter stability, and better prediction performance over a vast range of microwave and mmWave frequencies, distances, and scenarios, with fewer parameters~\cite{Sun16:TVT}.

In Version 1.4 of NYUSIM~\cite{NYUSIM_UM}, the PLE $n$ is set to the free space PLE of 2 with a shadow fading standard deviation of 4.0 dB for the line-of-sight (LOS) environment, while the PLE and shadow fading standard deviation are respectively set to 3.2 and 7.0 dB for the non-LOS (NLOS) environment~\cite{Rap15:TCOM,Samimi15:MTT}, for both UMi and UMa scenarios. For the RMa scenario, the LOS PLE and shadow fading standard deviation are 2.16 and 4.0 dB, respectively, while the NLOS PLE and shadow fading standard deviation are 2.75 and 8.0 dB, respectively~\cite{ACM,RMa}. All parameters may be adjusted by the user in MATLAB, as source code is provided using an open license~\cite{NYUSIM_UM}.


\subsection{Wideband Temporal/Spatial Clustering Algorithm}
The statistical spatial channel model (SSCM)~\cite{Samimi15:MTT} in NYUSIM utilizes \textit{time clusters} (TC) and \textit{spatial lobes} (SL) to model the omnidirectional CIR and corresponding joint angle of departure (AOD)/angle of arrival (AOA) power spectra, which have been used successfully in modeling mmWave channels~\cite{Samimi15:MTT}. \textit{Time clusters} are composed of multipath components (MPCs) traveling closely in time, and that arrive from potentially different angular directions in a short excess delay time window. \textit{Spatial lobes} represent main directions of arrival (or departure) where energy arrives over several hundreds of nanoseconds. This SSCM structure is motivated by field measurements, which have shown that multiple paths within a TC can arrive at unique pointing angles~\cite{Samimi15:MTT}, detectable due to high gain directional antennas, and this feature has not been modeled in current 3GPP and WINNER models~\cite{WinnerII,3GPP.36.873}. The TCSL approach implements a physically-based clustering scheme (e.g., the use of a fixed inter-cluster void interval representing the minimum propagation time between likely reflection or scattering objects) derived from field observations, and can be used to extract TC and SL statistics for any ray-tracing or measurement data sets~\cite{Samimi15:MTT}. The time-partitioning methodology delineates the beginning and end times of each time cluster, using a 25 ns minimum inter-cluster void interval. Sequentially arriving MPCs that occur within 25 ns of each other are assumed to belong to one TC. This is different than joint time-angle distributions~\cite{Ertel}. 

It is worth noting that in 3GPP TR 38.900 Release 14 for above 6 GHz~\cite{3GPP_Dec}, the number of clusters is unrealistically large. For example, in the UMi street canyon scenario, the number of clusters in the LOS environment is as high as 12, and 19 in the NLOS environment, which is not supported by the real-world measurements at mmWave bands~\cite{Rap13:Access,Rap15:TCOM,Samimi16:EuCAP,Samimi15:MTT}. In contrast, in the SSCM upon which NYUSIM is based~\cite{Samimi15:MTT}, the number of time clusters ranges from 1 to 6, and the mean number of spatial lobes is about 2 and is upper-bounded by 5, which are obtained from field observations and are much smaller than those in the 3GPP channel model~\cite{3GPP_Dec}. The impractically large number of clusters in the 3GPP channel model causes a higher rank of simulated mmWave channels, and unrealistic eigen-channel distributions, thereby yielding inaccurate spectral efficiency prediction for mmWave channels.

\section{Graphical User Interface and Simulator Basics}
\begin{figure*}
\centering
\captionsetup{justification=centering}
 \includegraphics[width=5.1in]{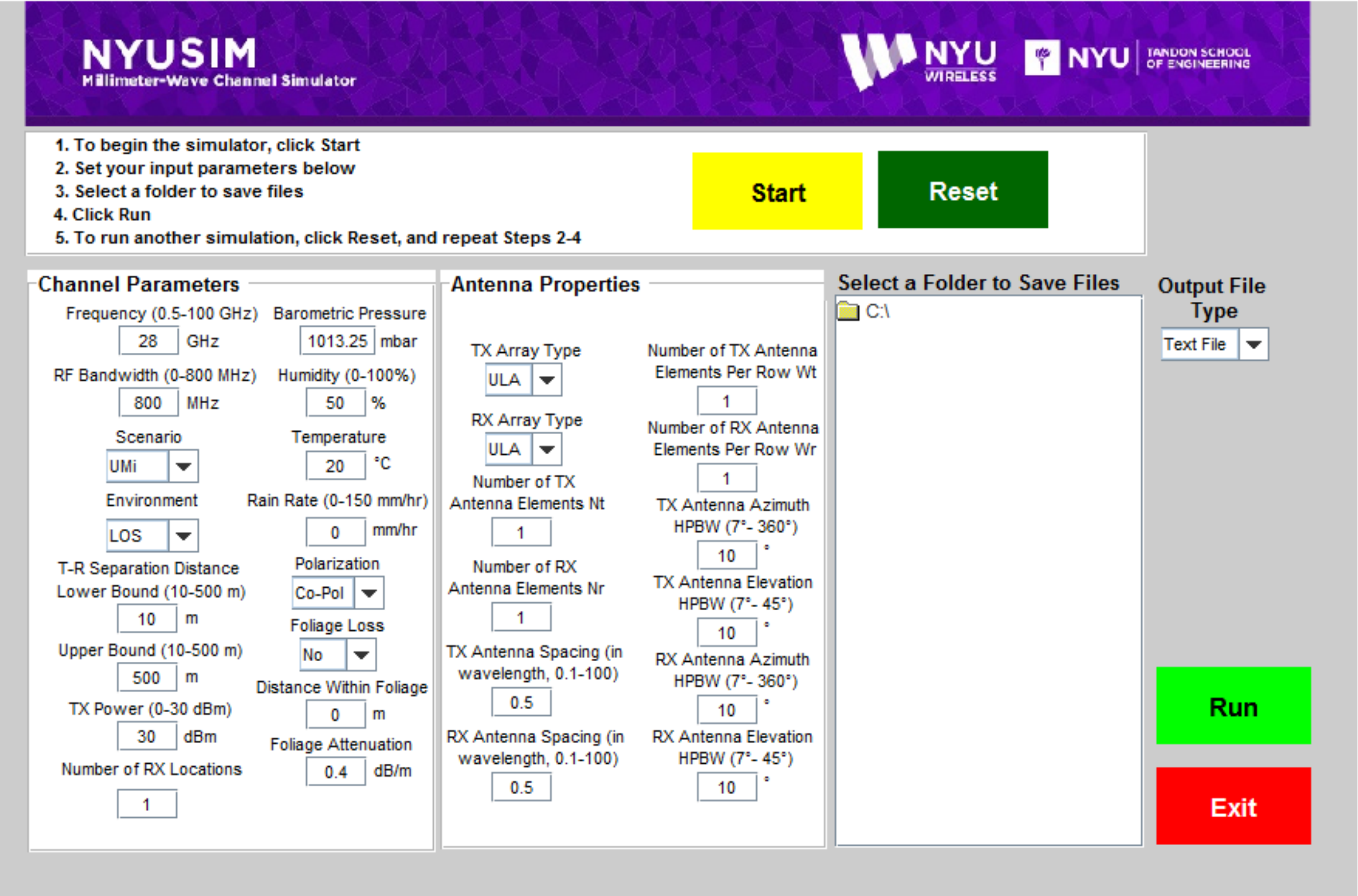}
    \caption{Graphical User Interface (GUI) of the NYUSIM channel model simulator.}
    \label{fig:GUI}
\end{figure*}

The screenshot in Fig.~\ref{fig:GUI} shows the graphical user interface (GUI) of NYUSIM. The simulator performs Monte Carlo simulations, generating certain numbers of samples of channel impulse responses (CIRs) at specific T-R separation distances, where the number of samples and the range of T-R separation distances are to be specified by users, as explained in the following subsection. It takes about 22 minutes to generate and save 100 CIRs and all the output files (five .png files, five sets of .txt files and five .mat files for each CIR simulation run, as well as one .png file, three sets of .txt files and three .mat files for the 100 simulation runs, as detailed in later subsections) on a PC server of Intel Xeon E5620 with two processors (2.40 GHz and 2.39 GHz) and 96.0 GB RAM~\cite{NYUSIM_UM}. NYUSIM runs about 30 times faster than our 3GPP MATLAB simulator.

\subsection{Input Parameters}
There are 28 input parameters to the simulator, which are grouped into two main categories: \textit{Channel Parameters} and \textit{Antenna Properties}, as shown on the GUI in Fig.~\ref{fig:GUI}. The panel \textit{Channel Parameters} contains 16 fundamental input parameters about the propagation channel, and the panel \textit{Antenna Properties} contains 12 input parameters related to the transmitter (TX) and receiver (RX) antenna arrays.

\subsection{Output Files}
\subsubsection{Output Figure Files}
\begin{figure}
	\centering
	\includegraphics[width=3.0in]{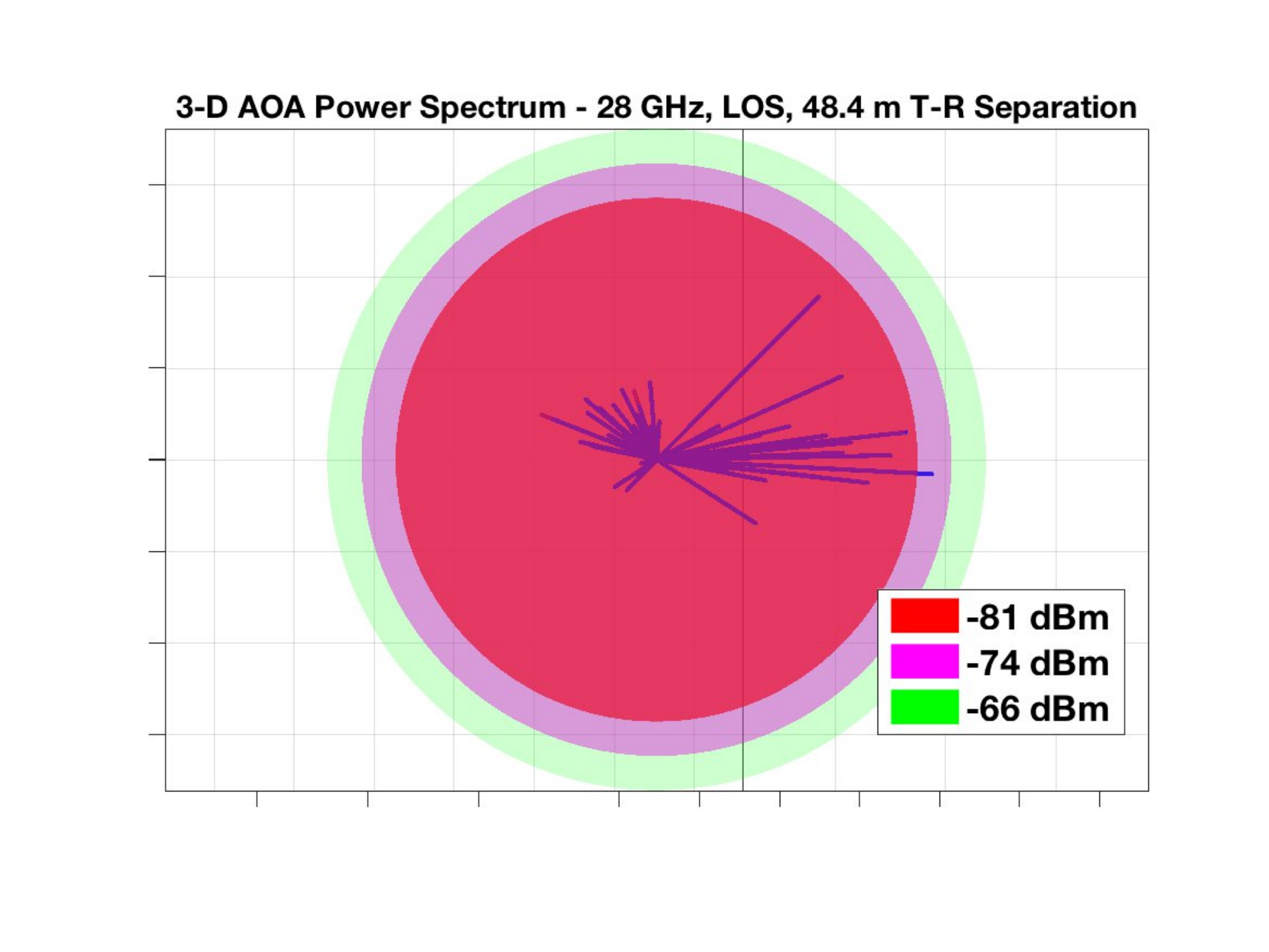}
	\caption{Example of a simulated AOA power spectrum.}
	\label{fig:AOA}
\end{figure}

\begin{figure}
\centering
 \includegraphics[width=3.0in]{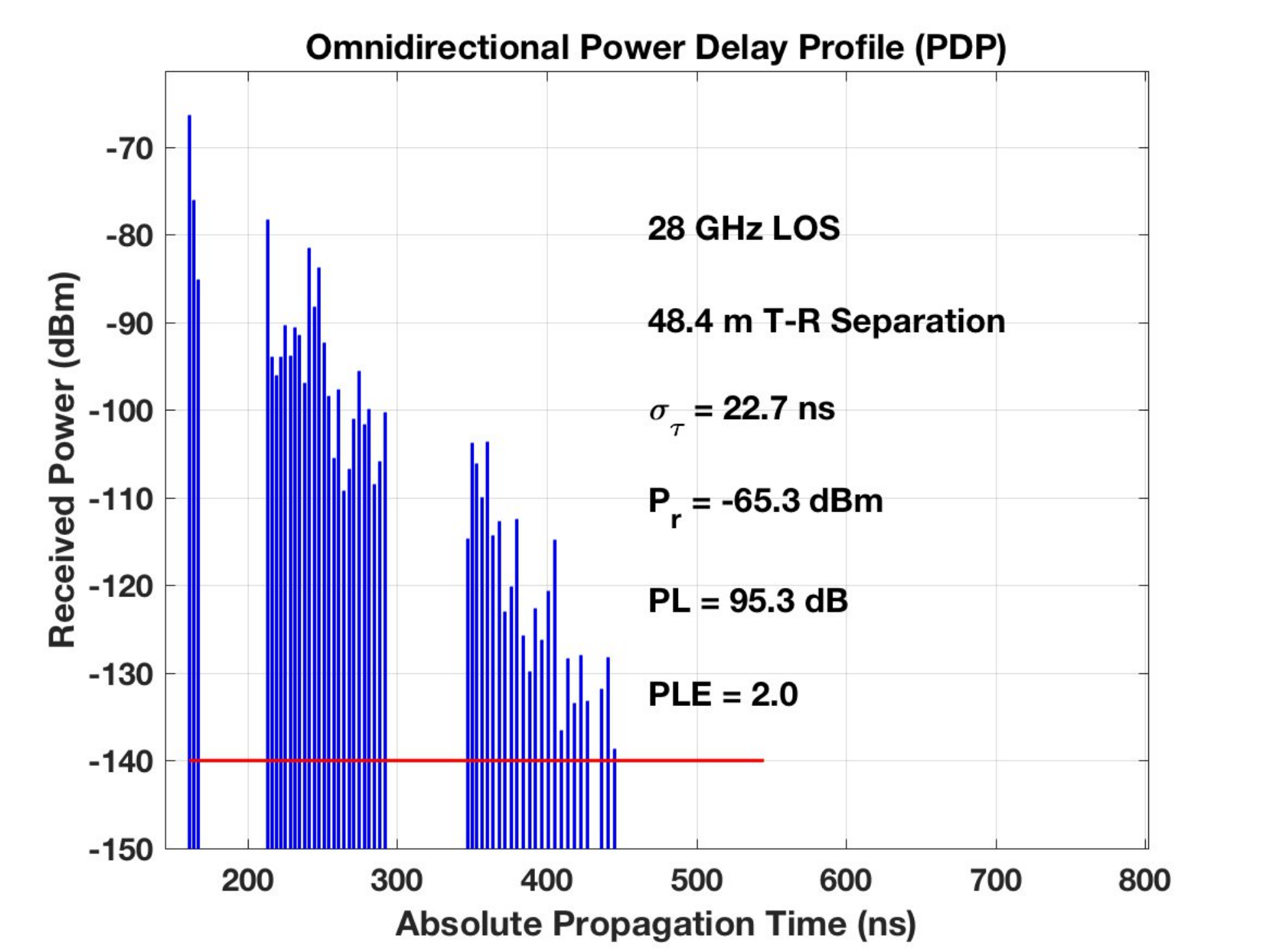}
    \caption{Example of a simulated omnidirectional PDP.}
    \label{fig:OmniPDP}
\end{figure}

For each simulation run, five figures will be generated and stored that are based on the particular results of the simulation that is being run, and an additional figure of path loss scatter plot will be generated and stored after $N~(N\geq1)$ continuous simulation runs with the same input parameters are complete. Regardless of the number of simulation runs (RX locations), the five figures generated from the first simulation run, as well as the last figure generated for $N$ continuous simulation runs, will pop up on the screen for visual purposes. The six output figures are: three-dimensional (3D) AOD power spectrum, 3D AOA power spectrum, a sample omnidirectional power delay profile (PDP), a sample directional PDP with strongest power, a series of PDPs over different receive antenna elements, and a path loss scatter plot. Fig.~\ref{fig:AOA} illustrates a simulated AOA power spectrum, with the corresponding simulated omnidirectional PDP depicted in Fig.~\ref{fig:OmniPDP}. 

\begin{figure}
	\centering
	\includegraphics[width=3.2in]{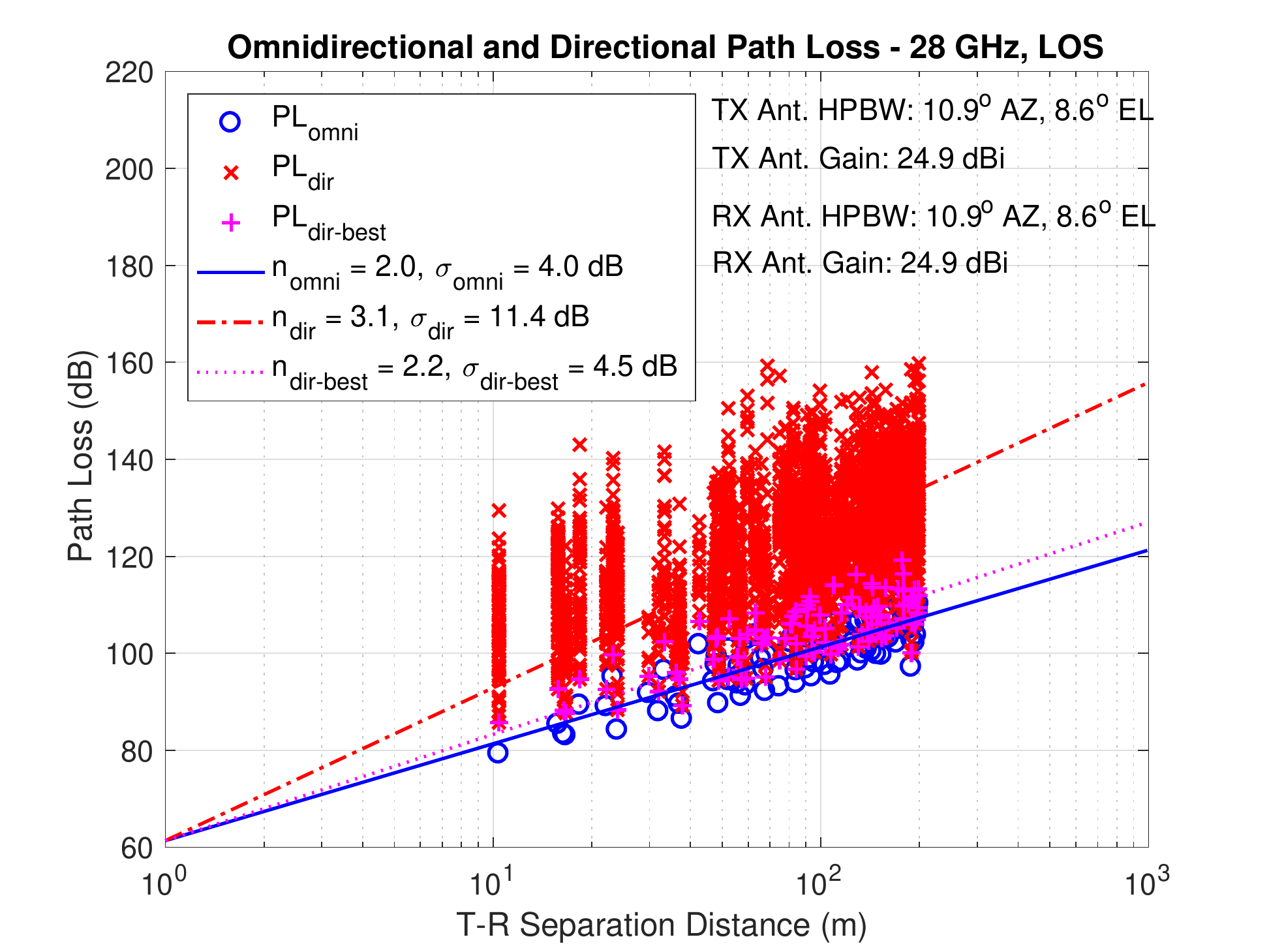}
	\caption{Example of a scatter plot showing the omnidirectional and directional path loss values generated from NYUSIM with 100 simulation runs.}
	\label{fig:PLScatter}
\end{figure}

Fig.~\ref{fig:PLScatter} displays the path loss scatter plot generated after $N$ continuous simulation runs, which shows omnidirectional path loss and directional path loss values for over the entire distance range generated from the $N$ continuous simulation runs, along with the fitted PLE and shadow fading standard deviation using the minimum-mean-square-error (MMSE) method~\cite{Sun16:TVT,Mac15:Indoor}. In the legend of Fig.~\ref{fig:PLScatter}, $n$ denotes the PLE, $\sigma$ is the shadow fading standard deviation, "omni" denotes omnidirectional, "dir" represents directional, and "dir-best" means the direction with the strongest received power. For producing the directional path loss at each RX location, NYUSIM searches for all possible pointing
angles in increments of the azimuth and elevation HPBWs of the TX/RX antenna specified by the user on the GUI after first generating the omnidirectional PDP. The TX/RX antenna gain pattern is calculated by NYUSIM using Eqs. (45) and (46) in~\cite{Samimi15:MTT} based on the azimuth and elevation HPBWs of TX and RX antennas specified by the user on the GUI. The directional path loss is equal to the transmit power plus the TX and RX antenna gains, minus the directional received power~\cite{Rap13:Access,Rap15:TCOM}. For generating Fig.~\ref{fig:PLScatter}, the antenna azimuth and elevation HPBWs are set to 10.9$^\circ$ and 8.6$^\circ$, respectively, at both the TX and the RX, to match the antenna HPBWs used in the 28 GHz measurements~\cite{Rap13:Access,Rap15:TCOM}. The simulated PLE and shadow fading standard deviation values agree well with the measured results presented in Table V and Table VIII of~\cite{Rap15:TCOM}. Directional path loss and directional PLE will always be larger (i.e., a directional channel is more lossy) than the omnidirectional case, because the directional antenna will spatially filter out many MPCs due to its directional pattern, such that the RX receives fewer MPCs hence less energy, thereby the directional path loss is higher after removing the antenna gain effect from the received power~\cite{Rap15:TCOM,Mac15:Indoor}. 

\subsubsection{Output Data Files}
For each simulation run, five sets of .txt files and five corresponding .mat files are generated, namely, \lq\lq{}AODLobePowerSpectrum$n$\_Lobe$x$.txt\rq\rq{},
\lq\lq{}AODLobePowerSpectrum$n$.mat\rq\rq{}, \lq\lq{}AOALobePowerSpectrum$n$\_Lobe$x$.txt\rq\rq{}, \lq\lq{}AOALobePowerSpectrum$n$.mat\rq\rq{}, \lq\lq{}OmniPDP$n$.txt", \lq\lq{}OmniPDP$n$.mat",
\lq\lq{}DirectionalPDP$n$.txt", \lq\lq{}DirectionalPDP$n$.mat", \lq\lq{}SmallScalePDP$n$.txt", and \lq\lq{}SmallScalePDP$n$.mat", where $n$ denotes the $n^{th}$ RX location (i.e., $n^{th}$ simulation run), and $x$ represents the $x^{th}$ spatial lobe. After $N$ continuous simulation runs, another three .txt files and three corresponding .mat files are produced, i.e., \lq\lq{}BasicParameters.txt", \lq\lq{}BasicParameters.mat", \lq\lq{}OmniPDPInfo.txt", \lq\lq{}OmniPDPInfo.mat", \lq\lq{}DirPDPInfo.txt", and \lq\lq{}DirPDPInfo.mat".

The files \lq\lq{}BasicParameters.txt" and \lq\lq{}BasicParameters.mat" subsume all the input parameter values as shown on the GUI when running the simulation. The files \lq\lq{}OmniPDPInfo.txt" and \lq\lq{}OmniPDPInfo.mat" contain four columns where each column represents a key parameter for each of the $N$ omnidirectional PDPs from $N$ continuous simulation runs, including T-R separation distance, omnidirectional received power, omnidirectional path loss, and omnidirectional root-mean-square (RMS) delay spread. The files \lq\lq{}DirPDPInfo.txt\rq\rq{} and \lq\lq{}DirPDPInfo.mat\rq\rq{} contain 10 columns where each column represents a key parameter for each of the directional PDPs from $N$ continuous simulation runs, such as time delay, received power, phase, azimuth and elevation AODs and AOAs of each resolvable MPC (i.e., antenna pointing angle), along with directional path loss and directional RMS delay spread. 

Each \lq\lq{}AODLobePowerSpectrum$n$\_Lobe$x$\rq\rq{} file is associated with a corresponding 3D AOD power spectrum output figure, and contains five parameters (columns) of each resolvable MPC in an AOD spatial lobe, namely: pathDelay (ns), pathPower (mWatts), pathPhase (rad), AOD (degree), and ZOD (degree). Similar parameters are contained in each .txt and .mat file \lq\lq{}AOALobePowerSpectrum$n$\_Lobe$x$\rq\rq{} that is associated with the output figure of 3D AOA power spectrum.

Each \lq\lq{}OmniPDP$n$\rq\rq{} file is associated with an omnidirectional PDP output figure, and contains two columns: the first column denotes the propagation time delay in nanoseconds, and the second column represents the received power in dBm. Each .txt and .mat file \lq\lq{}DirectionalPDP$n$\rq\rq{} is associated with the output figure of directional PDP with strongest power, and contains two columns: the first column denotes the propagation time delay in nanoseconds, and the second column represents the received power in dBm. Each .txt and .mat file \lq\lq{}SmallScalePDP$n$\rq\rq{} is associated with the output figure of the series of omnidirectional PDPs over different RX antenna elements, and contains three columns: the RX antenna separation in terms of number of wavelengths, the propagation time delay in nanoseconds, and the received power in dBm. 

\section{Applications of NYUSIM}
The output figures and data files generated from NYUSIM can be used in various ways based on users' needs, e.g., to simulate CIRs for mmWave systems, to investigate MIMO channel performance, and to perform BER simulation~\cite{Fung93,Thoma92}.

\subsection{MIMO Channel Condition Number}
First we show an example of how to obtain the condition number of a MIMO channel in NYUSIM by making use of the output data file \lq\lq{}DirPDPInfo.txt" or \lq\lq{}DirPDPInfo.mat", assuming orthogonal frequency division multiplexing (OFDM) modulation is utilized.

The condition number of a matrix is defined as the ratio of the largest to smallest singular value in the singular value decomposition of a matrix, and is a metric to characterize the quality of MIMO channels in the context of wireless communications~\cite{Rap15,Abb16,Heath05,Lu14}. The condition number will be high (e.g., over 20 dB) if the minimum singular value is close to zero, and will be 0 dB if singular values are equal. Physically, a small condition number value (e.g., below 20 dB) indicates good orthogonality of different sub-channels (a sub-channel usually has a distinct spatial direction), and the channel gains are comparable in different spatial directions. The rank of a matrix is the dimension of the vector space generated (or spanned) by its columns (or rows)~\cite{Bou:book}, and it determines how many data streams can be multiplexed over the channel in the context of MIMO communications~\cite{Abb16,Mat08,Sun14}. The condition number is related to the rank of a matrix: the matrix has full rank (the highest rank) when the condition number is equal or close to 0 dB (the lowest theoretical condition number). 

Now we explore the condition number of a MIMO channel matrix for a sub-carrier in an OFDM system. As described above, the output data files \lq\lq{}DirPDPInfo.txt" and \lq\lq{}DirPDPInfo.mat" contain paramount parameters of each resolvable MPC, which will be useful in generating the MIMO channel coefficient for an OFDM sub-carrier. Assuming uniform linear arrays (ULAs) are employed at both the TX and RX, the equation for generating such a channel coefficient is given by~\eqref{eq:hf}, which is adapted from Eq.(2) in~\cite{Adh14}, where $h_{m,k}(f)$ denotes the MIMO channel coefficient between the $m^{th}$ transmit antenna and the $k^{th}$ receive antenna for the sub-carrier $f$, $p$ represents the $p^{th}$ resolvable MPC, $\alpha$ is the amplitude of the channel gain, $\Phi$ denotes the phase of the MPC, $\tau$ represents the time delay, $d_T$ and $d_R$ are the antenna element spacing at the TX and RX, respectively, while $\phi$ and $\varphi$ denote the azimuth AOD and AOA, respectively. All of the above parameters can be extracted from the file "DirPDPInfo.txt" or "DirPDPInfo.mat". For each sub-carrier $f$ in a MIMO-OFDM system, there exists an $N_r\times N_t$ channel matrix \textbf{H} whose elements are $h_{m,k}(f)$, where $m=1,...,N_t$ and $k=1,...,N_r$. The condition number of \textbf{H} can be obtained consequently. The above mentioned approach along with the following input parameters settings on the NYUSIM GUI were used to run a simulation: 
\begin{itemize}
\item Frequency: 28 GHz
\item RF bandwidth: 800 MHz
\item Scenario: UMi
\item Environment: LOS
\item Lower Bound of T-R Separation Distance: 100 m
\item Upper Bound of T-R Separation Distance: 100 m
\item TX Power: 30 dBm
\item Barometric Pressure: 1013.25 mbar
\item Humidity: 50\%
\item Temperature: 20$^\circ$C
\item Rain Rate: 0 mm/hr
\item Polarization: Co-Pol
\item Foliage Loss: No
\item TX Array Type: ULA
\item RX Array Type: ULA
\item Number of TX Antenna Elements $N_t$: 2
\item Number of RX Antenna Elements $N_r$: 2
\item TX Antenna Spacing: 0.5 wavelength
\item RX Antenna Spacing: 0.5 wavelength
\item Number of TX Antenna Elements Per Row $W_t$: 2
\item Number of RX Antenna Elements Per Row $W_r$: 2
\item TX Antenna Azimuth HPBW: 10$^\circ$
\item TX Antenna Elevation HPBW: 10$^\circ$
\item RX Antenna Azimuth HPBW: 10$^\circ$
\item RX Antenna Elevation HPBW: 10$^\circ$
\end{itemize}
\begin{figure*}
	\begin{equation}\label{eq:hf}
	h_{m,k}(f)=\sum_{p}\alpha_{m,k,p}e^{j\Phi_{m,k,p}}e^{-j2\pi f\tau_{m,k,p}}e^{-j2\pi d_{T}m\sin(\phi_{m,k,p})}e^{-j2\pi d_{R}k\sin(\varphi_{m,k,p})}
	\end{equation}
\end{figure*}

\begin{figure}
\centering
 \includegraphics[width=2.5in]{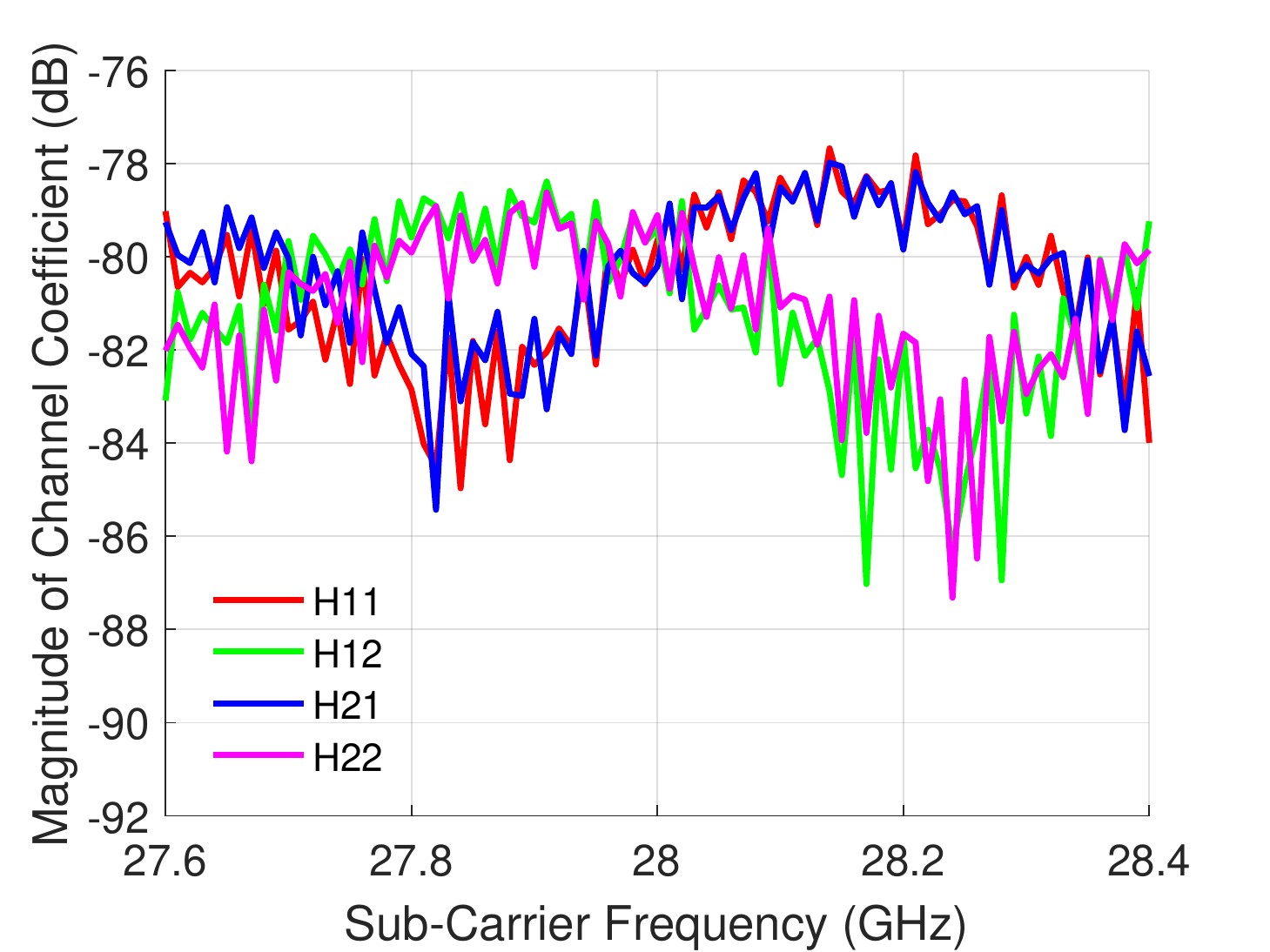}
    \caption{MIMO channel matrix coefficient magnitudes for OFDM sub-carriers with two transmit and two receive antenna elements for MIMO-OFDM channels in one simulation run.}
    \label{fig:HChannelCoef}
\end{figure}

For the simulation, assuming the frequency interval between adjacent sub-carriers is 10 MHz, which corresponds to 800 MHz/10 MHz + 1 = 81 sub-carriers, we performed a simulation to obtain an intuitive observation on the MIMO channel matrix coefficients and condition number of the channel matrix. Fig.~\ref{fig:HChannelCoef} illustrates the resultant magnitude of the four channel coefficients, wherein the transmitted wideband signal undergoes frequency-selective fading, and the fading magnitude varies for different TX-RX antenna element combinations. 


Next, with the identical input parameters listed above, but the frequency interval between adjacent sub-carriers set to 500 kHz, which corresponds to 800 MHz/500 kHz + 1 = 1601 sub-carriers, we performed a simulation to emulate a random MIMO channel realization with the input parameters described above. Then the following changes are made to the four input parameters below with all the other input parameter values remaining the same:
\begin{itemize}
\item Number of TX Antenna Elements $N_t$: 3
\item Number of RX Antenna Elements $N_r$: 3
\item Number of TX Antenna Elements Per Row $W_t$: 3
\item Number of RX Antenna Elements Per Row $W_r$: 3
\end{itemize}

\begin{figure}
\centering
 \includegraphics[width=2.4in]{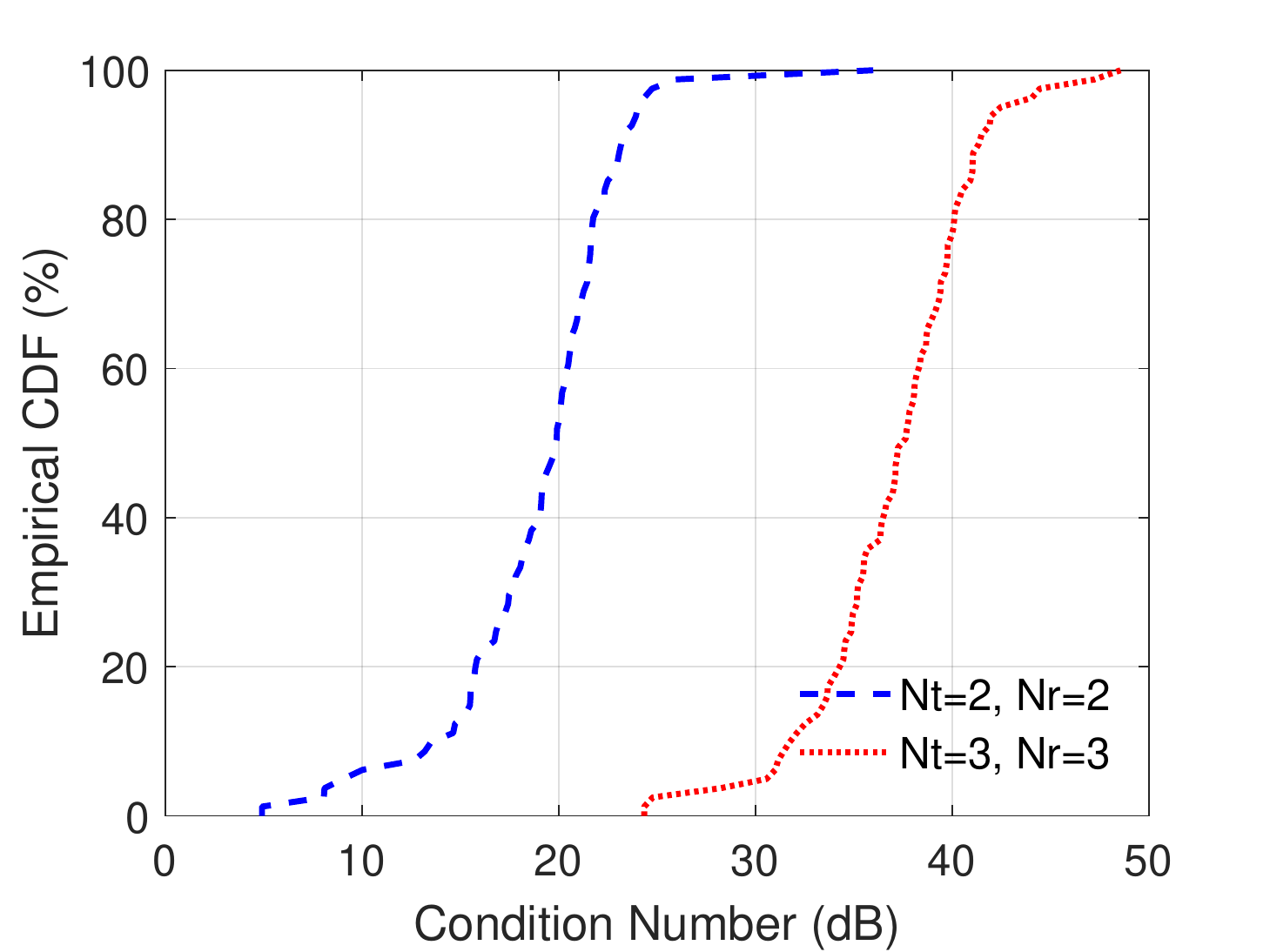}
    \caption{CDF of the condition number of channel matrices for OFDM sub-carriers with different transmit and receive antenna elements $N_t$ and $N_r$ for MIMO-OFDM channels in one simulation run.}
    \label{fig:CondCDF1}
\end{figure}

Fig.~\ref{fig:CondCDF1} depicts the empirical cumulative distribution function (CDF) of the condition number of channel matrices for OFDM sub-carriers with the above two sets of input parameters from the simulation. It is apparent from Fig.~\ref{fig:CondCDF1} that the condition numbers of the individual OFDM sub-carriers for a $3\times 3$ MIMO channel is about 18 dB larger compared to the $2\times 2$ case on average, and the relatively large condition number of the $3\times 3$ channel matrix may stem from the fact that the matrix is rank deficient. 

\subsection{Spectral Efficiency Comparison Between 3GPP and NYUSIM Channel Models}
As mentioned in Section II-B, the 3GPP TR 38.900 Release 14 channel model~\cite{3GPP_Dec} contains unrealistic number of clusters (e.g., 19 clusters for UMi NLOS) and up to 20 rays per cluster, which is excessively large and not borne out by measurements. In this subsection, we use the 3GPP TR 38.900 Release 14 channel model~\cite{3GPP_Dec} and NYUSIM channel model to analyze and compare the spectral efficiency (SE) for mmWave MIMO channels.

Let us assume a single-cell single-user MIMO system operating at 28 GHz with an RF bandwidth of 100 MHz in the UMi scenario. The base station is equipped with 256 antenna elements comprising a uniform rectangular array (URA), where the antenna pattern on Page 23 of~\cite{3GPP_Dec} is adopted. The user has 16 antenna elements constituting a URA. Signal-to-noise ratios (SNRs) are fixed at certain values to investigate the SE achieved by the 3GPP~\cite{3GPP_Dec} and NYUSIM channel models. Two hundred random channel realizations with distances ranging from 10 to 435 m were performed for each channel model. 

\begin{figure}
	\centering
	\includegraphics[width=2.5in]{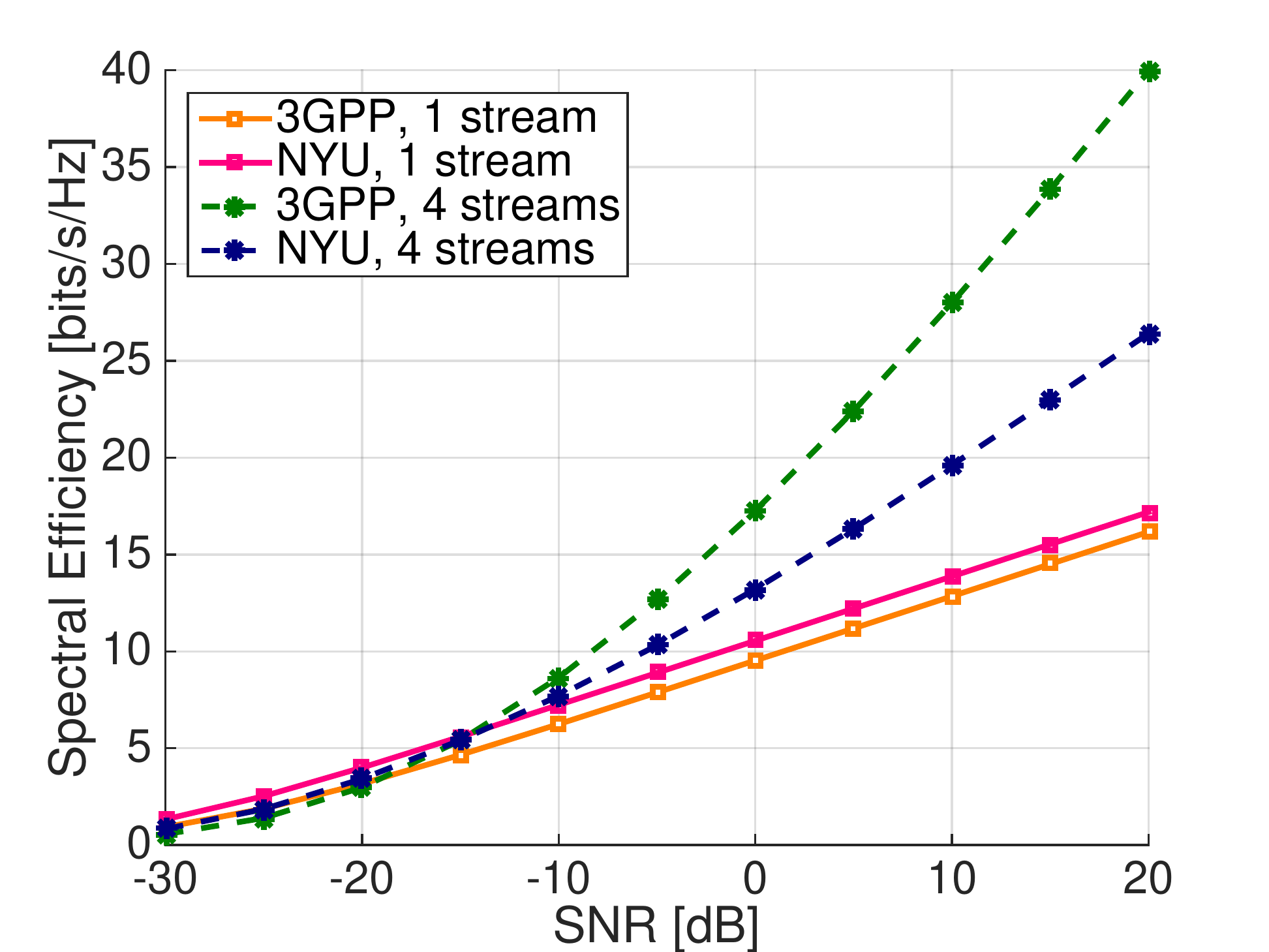}
	\caption{Spectral efficiency achieved by hybrid beamforming for a 256$\times$16 system at 28 GHz with rectangular antenna arrays at the transmitter and receiver with four RF chains. The results were averaged over 200 simulation runs.}
	\label{fig:SE1}
\end{figure}

Fig.~\ref{fig:SE1} illustrates the SE achieved by the hybrid beamforming algorithm proposed in~\cite{Ayach} for a 256$\times$16 mmWave system at 28 GHz for various numbers of data streams $N_s$, using the 3GPP and NYUSIM models. As shown by Fig.~\ref{fig:SE1}, the SE generated by the 3GPP channel model is just slightly smaller than that yielded by the NYUSIM channel model for $N_s=1$, but is much larger for $N_s=4$. For instance, for an SNR of 20 dB and $N_s=4$, the 3GPP SE (40 bits/s/Hz) is about 13 bits/s/Hz greater than the NYUSIM SE (about 27 bits/s/Hz). This is because NYUSIM yields one or two strong dominant clusters and much weaker non-dominant clusters, while the 3GPP model has less focused directional energy than what realistically exists. The results indicate that the 3GPP channel model is optimistic when predicting diversity and the achievable SE at mmWave frequencies, while NYUSIM provides realistic channel parameters and SE predictions due to the use of extensive real-world measurement data at mmWave frequencies. NYUSIM can help avoid system errors inherent with legacy modeling approaches.

\section{Conclusion}
This paper has presented an open-source channel software simulator, NYUSIM, developed from extensive broadband propagation measurements at mmWave frequencies. NYUSIM recreates wideband PDPs/CIRs and channel statistics for a variety of carrier frequencies, RF bandwidths, antenna beamwidths, environment scenarios, and atmospheric conditions, and is equipped with a GUI that makes the simulator more user-friendly. Over 7000 downloads have already been logged by major corporations and universities worldwide. Simulated results from NYUSIM match well with the measured data. Application examples study channel conditions and compare MIMO channel performance between 3GPP and NYUSIM models. NYUSIM can be employed to perform various other types of analysis and is useful for 5G communication system development and deployment. 


\ifCLASSOPTIONcaptionsoff
  \newpage
\fi

\bibliographystyle{IEEEtran}
\bibliography{bibliography_NYU}

\end{document}